\documentclass[12pt]{article}
\usepackage{epsf}
\usepackage[dvips]{graphicx}
\usepackage{amssymb}

\begin{document}
\title{Inclusive cross sections of proton-proton and
proton-antiproton scattering}

\author{{\slshape Victor Abramovsky, Natalia Radchenko}\\[1ex]
Novgorod State University, 173003 Novgorod the Great, Russia}

\maketitle

\begin{abstract}
We have predicted the difference in inclusive cross sections on
pseudorapidity in $pp$ and $p\bar{p}$ interactions at
$\sqrt{s}=900$~GeV. Their ratio $R=\left({\rm
d}\sigma^{p\bar{p}}/{\rm d}\eta\right)\left/\left({\rm
d}\sigma^{pp}/{\rm d}\eta\right)\right.>1$ in the whole
pseudorapidity range. On the basis of AGK theorem we show that the
ratio of inclusive cross sections of $pp$ and $p\bar{p}$ at
$\sqrt{s}=900$~GeV in the region of low transverse momenta
$p_\perp$ up to 2~GeV $\left(\frac{1}{2\pi p_\perp}\frac{{\rm
d}^2\sigma^{p\bar{p}}}{{\rm d}\eta {\rm
d}p_\perp}\right)\left/\left(\frac{1}{2\pi p_\perp}\frac{{\rm
d}^2\sigma^{pp}}{{\rm d}\eta {\rm d}p_\perp}\right)\right.=R$.
Experimental measurements by the ATLAS Coll. give value
$R\simeq1.2$ for interval $|\eta|<2.5$.

The difference in inclusive cross sections results from presence
of additional subprocess in $p\bar{p}$ --  hadrons production from
decay of three quark strings, which is absent in $pp$ scattering.
\end{abstract}

\section{Introduction}
It is generally accepted that total cross sections of $pp$ and
$p\bar{p}$ interactions at high energies are the same
$$
\sigma_{tot}^{pp}\equiv\sigma_{tot}^{p\bar{p}}\quad\quad\quad(s\rightarrow\infty).
$$
Also elastic cross sections are equal
$\sigma_{el}^{pp}\equiv\sigma_{el}^{p\bar{p}}$ and differential
elastic cross sections are equal ${\rm d}\sigma_{el}^{pp}/{\rm
d}t\equiv {\rm d}\sigma_{el}^{p\bar{p}}/{\rm d}t$.

It follows from the Pomeranchuk theorem~\cite{bib1}, which was
proved for constant total cross sections in condition
$s\rightarrow\infty$. This theorem  was generalized by
Iden~\cite{bib2} for increasing total cross sections, which
fulfill the Froissare theorem.

It is also generally accepted that multiplicity properties of $pp$
and $p\bar{p}$ interactions, such as inclusive cross sections and
distributions of charged particles are the same.

In the paper~\cite{Abramovsky:2009ni} we have shown that inclusive
cross sections of $p\bar{p}$ scattering ${\rm
d}\sigma^{p\bar{p}}/{\rm d}\eta$ are larger than inclusive cross
sections of $pp$ scattering ${\rm d}\sigma^{pp}/{\rm d}\eta$ at
$\sqrt{s}=900$~GeV.

The ATLAS Coll. have published data~\cite{Aad:2010rd} on inclusive
cross sections from which it follows that
\begin{equation}\label{1}
\frac{1}{2\pi p_\perp}\frac{{\rm d}^2\sigma^{p\bar{p}}}{{\rm
d}\eta {\rm d}p_\perp}\left/\frac{1}{2\pi p_\perp}\frac{{\rm
d}^2\sigma^{pp}}{{\rm d}\eta {\rm d}p_\perp}\right.\simeq1.2
\end{equation}
for pseudorapidity $|\eta|<2.5$ in interval of transverse momenta
up to $p_\perp\simeq2$~GeV/c.

The ALICE Coll. confirmed this result~\cite{Aamodt:2010my} for
$|\eta|<0.8$ and transverse momenta up to
$p_\perp\simeq1.5$~GeV/c.

In the present work we will show that ratio~(\ref{1}) is equal to
\begin{equation}\label{2}
\frac{1}{2\pi p_\perp}\frac{{\rm d}^2\sigma^{p\bar{p}}}{{\rm
d}\eta {\rm d}p_\perp}\left/\frac{1}{2\pi p_\perp}\frac{{\rm
d}^2\sigma^{pp}}{{\rm d}\eta {\rm d}p_\perp}\right.=\frac{{\rm
d}\sigma^{p\bar{p}}}{{\rm d}\eta}\left/\frac{{\rm
d}\sigma^{pp}}{{\rm d}\eta}\right.=R.
\end{equation}

We have estimated the value $R\simeq1.12$ both for pseudorapidity
intervals $|\eta|<0.8$ and  $|\eta|<2.5$.

\section{Low Constituents Number Model}
We are based on the Low Constituents Number Model
(LCNM)~\cite{Abramovsky:1980yc}, \cite{Abramovsky:1980kk},
\cite{bib9} which can be represented as follows.

1. On the first step before the collision there is small number of
constituents in initial hadrons. In every hadron there are
components either with only valence quarks or with valence quarks
and one additional gluon.

2. On the second step the hadrons interaction is carried out by
gluon exchange between the valence quarks and initial gluons. The
hadrons gain the color charge.

3. On the third step after interaction the colored hadrons move
apart and when the distance between them becomes larger than the
confinement radius, the lines of color electric field gather into
the string. This string breaks out into secondary hadrons.

Hadrons production in this model is depicted by phenomenological
diagrams in Fig.~1. Solid lines correspond to valence quarks and
antiquarks. Wavy line corresponds to exchanging gluon, which
performs the interaction. Additional gluons are also represented
by wavy lines, one in every projectile hadron.  Formation of color
field string (shown as spiral) and its breakup to secondary
hadrons takes place at the third stage. This process corresponds
to interaction in final state. It should be noted that there is
subprocesses of hadrons production from three quark strings in
$p\bar{p}$ interaction. There is no such subprocess in $pp$
interaction. This subprocess defines the difference in inclusive
cross sections and multiplicity distributions in $pp$ and
$p\bar{p}$.

\begin{figure}[!h]
\centerline{
\includegraphics[scale=0.5]{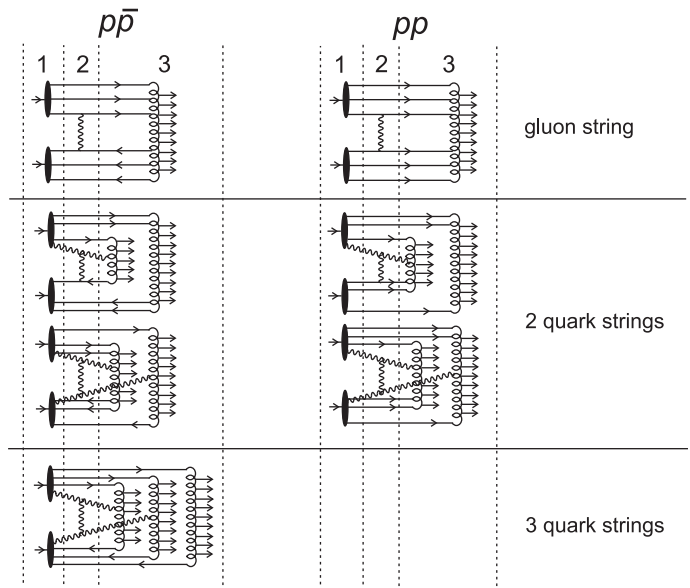}}
\caption{Three types of inelastic subprocesses in $pp$ and
$p\bar{p}$ scattering in the LCNM.}
\end{figure}

We consider only one gluon exchange, this is our hypothesis.
Therefore interaction between components with only valence quarks
both in $pp$ and $p\bar{p}$ leads to separation of color charges
with octet quantum numbers.

Since we consider only one gluon exchange, there is no three-sheet
configuration which was proposed in~\cite{Rossi:1977cy},
\cite{bib11}  in the diagram with only valence quarks in proton
and valence antiquarks in antiproton.

\section{Difference in inclusive cross sections of $pp$ and $p\bar{p}$}

Let us define topological inclusive cross section of production of
one charged particle in case with $n$ charged particles

\begin{equation}\label{3}
(2\pi)^32E\frac{{\rm d}^3\sigma_n^{incl}}{{\rm
d}^3p},\;\;\;\;\int{\rm d}^3p\frac{{\rm d}^3\sigma_n^{incl}}{{\rm
d}^3p}=n\,\sigma_n,
\end{equation}
where $\sigma_n$ -- topological cross sections of $n$ charged
particles production. We consider only non single diffractive
events so
\begin{equation}\label{4}
\sum\sigma_n=\sigma^{nsd}.
\end{equation}

Here we are based on the UA5 Coll. data~\cite{Alner:1986xu} on the
inclusive cross sections in 9 multiplicity bins: $2\leqslant
n\leqslant 10$, $12\leqslant n \leqslant20$, \ldots, $n\geqslant
82$. In accordance with UA5 Coll. we define inclusive cross
sections in every of these bins ($i=1,\ldots,9$)
\begin{equation}\label{5}
\frac{{\rm d}^3\sigma^{(i)incl}}{{\rm
d}^3p}=\sum_{n\mbox{\scriptsize{ in bin }}(i)}\!\!\!\frac{{\rm
d}^3\sigma_n^{incl}}{{\rm d}^3p},\;\;\;\;\left(\frac{{\rm
d}^3\sigma^{incl}}{{\rm d}^3p}=\sum_{i=1}^9\frac{{\rm
d}^3\sigma^{(i)incl}}{{\rm d}^3p}\right),
\end{equation}
which are normalized as follows
\begin{equation}\label{6}
\int{\rm d}^3p\frac{{\rm d}^3\sigma^{(i)incl}}{{\rm
d}^3p}=\sum_{n\mbox{\scriptsize{ in bin
}}(i)}\!\!\!n\,\sigma_n=\sigma^{nsd}\sum_{n\mbox{\scriptsize{ in
bin }}(i)}\!\!\!n\,P_n=\sigma^{nsd}\,\bar{n}^{(i)}
\end{equation}
where $P_n=\sigma_n/\sigma^{nsd}$ -- probability of $n$ charged
particles production in non single diffractive event. Inasmuch as
we consider that inclusive cross sections of $pp$ and $p\bar{p}$
are different we write down relation~(\ref{6}) separately for $pp$
and $p\bar{p}$
\begin{equation}\label{7}
\int{\rm d}^3p\frac{{\rm d}^3\sigma^{(i)incl}_{p\bar{p}}}{{\rm
d}^3p}=\sigma^{nsd}\,\bar{n}^{(i)}_{p\bar{p}},\;\;\;\; \int{\rm
d}^3p\frac{{\rm d}^3\sigma^{(i)incl}_{pp}}{{\rm
d}^3p}=\sigma^{nsd}\,\bar{n}^{(i)}_{pp}.
\end{equation}

It was shown in~\cite{Abramovsky:2009hd} that $\sigma^{sd}$ --
cross section of single diffractive events is the same for $pp$
and $p\bar{p}$ interactions at high energies. Therefore cross
section $\sigma^{nsd}=\sigma_{tot}-\sigma_{el}-\sigma_{sd}$ is
also the same for $pp$ and $p\bar{p}$ interactions.

From ratio of $p\bar{p}$ over $pp$ in (\ref{7}) we obtain the
following relation
\begin{equation}\label{8}
\int{\rm d}^3p\frac{{\rm d}^3\sigma^{(i)incl}_{p\bar{p}}}{{\rm
d}^3p}=\frac{\bar{n}^{(i)}_{p\bar{p}}}{\bar{n}^{(i)}_{pp}}\int{\rm
d}^3p\frac{{\rm d}^3\sigma^{(i)incl}_{pp}}{{\rm d}^3p}.
\end{equation}
Value of $\bar{n}^{(i)}_{p\bar{p}}/\bar{n}^{(i)}_{pp}$ does not
depend on momentum of observed particle $p$. Therefore one of
solutions of (\ref{8}) (perhaps, the only solution) has the form
\begin{equation}\label{9}
\frac{{\rm d}^3\sigma^{(i)incl}_{p\bar{p}}}{{\rm
d}^3p}=\frac{\bar{n}^{(i)}_{p\bar{p}}}{\bar{n}^{(i)}_{pp}}\,\frac{{\rm
d}^3\sigma^{(i)incl}_{pp}}{{\rm d}^3p}.
\end{equation}
(If $pp$ and $p\bar{p}$ interactions are the same then we obtain a
trivial result.)

Factorization of inclusive cross sections results from the AGK
theorem~\cite{Abramovsky:1973fm}, so we can write down
\begin{equation}\label{10}
\frac{1}{2\pi p_\perp}\frac{{\rm d}^2\sigma^{incl}_{pp}}{{\rm
d}\eta{\rm d}p_\perp}=f_{pp}(p_\perp)\frac{{\rm
d}\sigma^{incl}_{pp}}{{\rm d}\eta},\;\;\;\; \frac{1}{2\pi
p_\perp}\frac{{\rm d}^2\sigma^{incl}_{p\bar{p}}}{{\rm d}\eta{\rm
d}p_\perp}=f_{p\bar{p}}(p_\perp)\frac{{\rm
d}\sigma^{incl}_{p\bar{p}}}{{\rm d}\eta}
\end{equation}
where ${\rm d}\sigma^{incl}/{\rm d}\eta$ -- inclusive distribution
on pseudorapidity. It is easy to show from the equation~(\ref{9})
that
\begin{equation}\label{11}
f_{pp}(p_\perp)=f_{p\bar{p}}(p_\perp).
\end{equation}
Therefore from ratio of $p\bar{p}$ over $pp$ in (\ref{10}) we can
obtain strict equality
\begin{equation}\label{12}
\left(\frac{1}{2\pi p_\perp}\frac{{\rm
d}^2\sigma^{incl}_{p\bar{p}}}{{\rm d}\eta{\rm
d}p_\perp}\right)\left/\left(\frac{1}{2\pi p_\perp}\frac{{\rm
d}^2\sigma^{incl}_{pp}}{{\rm d}\eta{\rm
d}p_\perp}\right)\right.=\left(\frac{{\rm
d}\sigma^{incl}_{p\bar{p}}}{{\rm
d}\eta}\right)\left/\left(\frac{{\rm d}\sigma^{incl}_{pp}}{{\rm
d}\eta}\right)\right..
\end{equation}

In the paper~\cite{Abramovsky:2009ni} we have obtained the
inclusive cross sections ${\rm d}\sigma^{incl}_{pp}/{\rm d}\eta$
in $pp$ interactions at $\sqrt{s}=900$~GeV. The result is shown in
Fig.~2 together with experimental data of the UA5
Coll.~\cite{Alner:1986xu}. From this graph it can be seen that the
ratio $R=\left({\rm d}\sigma^{incl}_{p\bar{p}}/{\rm
d}\eta\right)\left/\left({\rm d}\sigma^{incl}_{pp}/{\rm
d}\eta\right)\right.\simeq1.12$  both for pseudorapidity intervals
$|\eta|<0.8$ and  $|\eta|<2.5$.

\begin{figure}[!h]
\centerline{
\includegraphics[scale=0.6]{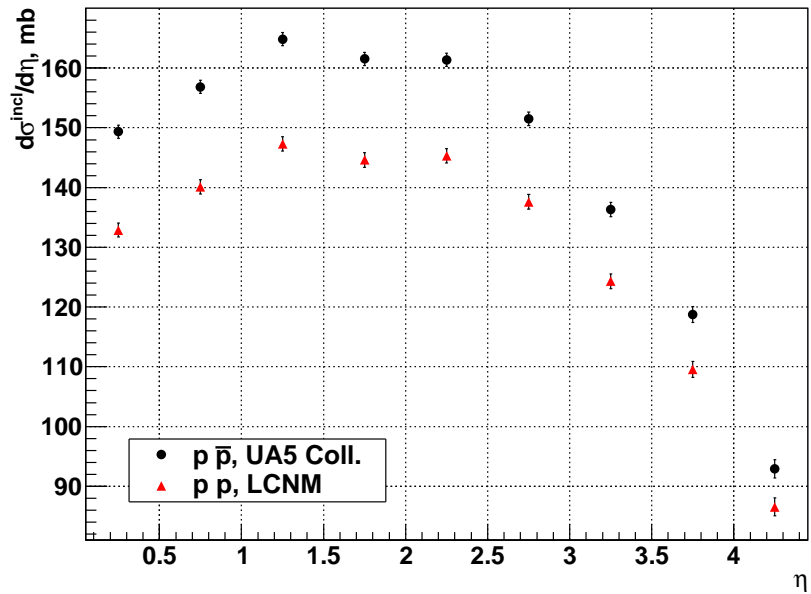}}
\caption{Absolute value of inclusive cross section at
$\sqrt{s}=900$~GeV~\cite{Abramovsky:2009ni}.}
\end{figure}

It should be noted that the relation~(\ref{10}) must be fulfilled
in region of soft physics for transverse momenta up to
$p_\perp=1.5\div2$~GeV/c, where the AGK theorem is valid.
Therefore, in this region the ratio of inclusive cross sections
\begin{equation}\label{13}
\left(\frac{1}{2\pi p_\perp}\frac{{\rm
d}^2\sigma^{incl}_{p\bar{p}}}{{\rm d}\eta{\rm
d}p_\perp}\right)\left/\left(\frac{1}{2\pi p_\perp}\frac{{\rm
d}^2\sigma^{incl}_{pp}}{{\rm d}\eta{\rm
d}p_\perp}\right)\right.=R\simeq1.12.
\end{equation}

The comparison of experimental results of collaborations ATLAS and
ALICE for $pp$ with results of the UA1 Coll.~\cite{Albajar:1989an}
for $p\bar{p}$ is shown in Fig.~3 together with our estimation
(graphs are taken from \cite{Aad:2010rd} and
\cite{Aamodt:2010my}). In our opinion, these data show that
inclusive cross sections of $p\bar{p}$ interaction exceed
inclusive cross sections of $pp$ interaction. We suppose that the
excess is determined by presence of subprocess of hadrons
production from three quark strings, which exists in $p\bar{p}$
and is absent in $pp$ interactions.

\begin{figure}[!h]
\centerline{
\includegraphics[scale=0.6]{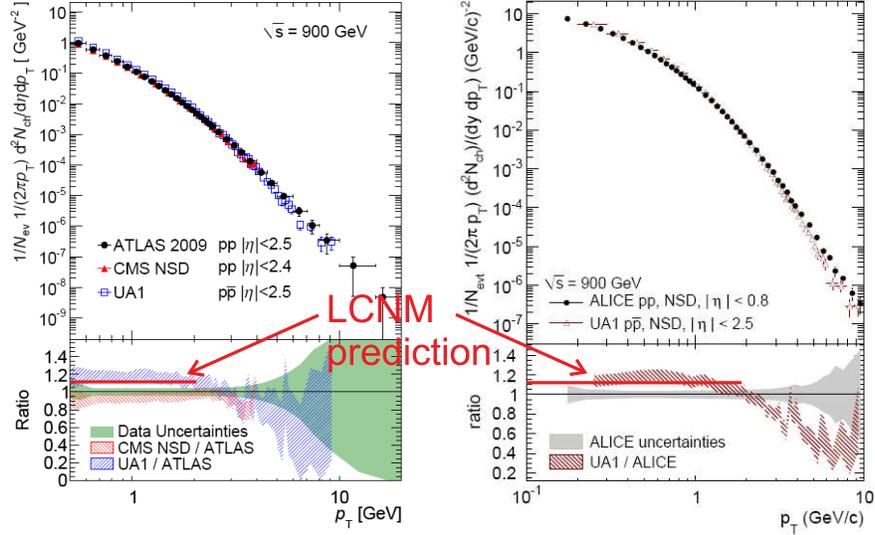}}
\caption{Ratio of inclusive cross sections of $p\bar{p}$ (UA1
Coll.~\cite{Albajar:1989an}) over $pp$ (ATLAS
Coll.~\cite{Aad:2010rd} and ALICE Coll.~\cite{Aamodt:2010my}) and
Low Constituents Number Model prediction.}
\end{figure}

We will not comment here the explanation of difference of
inclusive cross sections ratio from 1 by systematic uncertainties
of the experiments UA1, ATLAS and ALICE. This question demands
thorough analysis.

It should be noted that the presence of inelastic subprocess of
hadrons production from three quark strings might be important in
nucleus-nucleus collisions. Protons and antiprotons produced from
the first collisions will give different cascades when passing
through nucleus.

\section{Conclusion}
We think that ATLAS Coll. discovered a new effect -- the
difference in multiparticle production in $pp$ and $p\bar{p}$
interactions. ALICE Coll. confirmed this effect. We propose to
experimentalists to measure inclusive cross section in bins with
high multiplicities, where the difference will be the most
evident.

On the basis of the Low Constituents Number Model we also predict
the value
$\displaystyle\left.\frac{d\sigma^{pp}}{d\eta}\right|_{\eta=0}=6.62\pm0.70$
at energy $\sqrt{s}=14$~TeV.

\section*{Acknowledgements}
We thank O.~V.~Kancheli for discussion and critical remarks which
led to better understanding of the problem.

Victor Abramovsky gratefully acknowledges financial support from
the Russian Foundation for Basic Research, grant 10-02-68536-z.

Natalia Radchenko gratefully acknowledges financial support by
grant of Ministry of education and science of the Russian
Federation, federal target program ``Scientific and
scientific-pedagogical personnel of innovative Russia'', grant
P1200.


\begin{thebibliography}{99}

\bibitem{bib1}
I.~Ya.~Pomeranchuk, Zh.\ Eksp.\ Teor.\ Fiz.\ \textbf{34} (1958)
725.

\bibitem{bib2}
R.~J.~Iden. High Energy Collisions of Elementary Particles,
Cambridge U.P., 1967 -- 309 p.

\bibitem{Abramovsky:2009ni}
  V.~A.~Abramovsky and N.~V.~Radchenko,
  arXiv:0912.1041 [hep-ph].


\bibitem{Aad:2010rd}
  G.~Aad {\it et al.}  [ATLAS Collaboration],
  Phys.\ Lett.\  B {\bf 688} (2010) 21
  [arXiv:1003.3124 [hep-ex]].


\bibitem{Aamodt:2010my}
  K.~Aamodt {\it et al.}  [ALICE Collaboration],
  Phys.\ Lett.\  B {\bf 693} (2010) 53
  [arXiv:1007.0719 [hep-ex]].

\bibitem{Abramovsky:1980yc}
  V.~A.~Abramovsky and O.~V.~Kancheli,
  Pisma Zh.\ Eksp.\ Teor.\ Fiz.\  {\bf 31} (1980) 566.


\bibitem{Abramovsky:1980kk}
  V.~A.~Abramovsky and O.~V.~Kancheli,
  Pisma Zh.\ Eksp.\ Teor.\ Fiz.\  {\bf 32} (1980) 498.

\bibitem{bib9}
V. A. Abramovsky and N. V. Radchenko,   Particles and Nuclei,
Letters \textbf{6} (2009) 607 [arXiv:0812.2465 [hep-ph]].

\bibitem{Rossi:1977cy}
  G.~C.~Rossi and G.~Veneziano,
  Nucl.\ Phys.\  B {\bf 123} (1977) 507.


\bibitem{bib11}
B. Z. Kopeliovich and B. G. Zakharov, preprint Dubna E2-87-911
(1987).

\bibitem{Alner:1986xu}
  G.~J.~Alner {\it et al.}  [UA5 Collaboration],
  Z.\ Phys.\  C {\bf 33} (1986) 1.


\bibitem{Abramovsky:2009hd}
  V.~A.~Abramovsky,
  arXiv:0911.4850 [hep-ph].

\bibitem{Abramovsky:1973fm}
  V.~A.~Abramovsky, V.~N.~Gribov and O.~V.~Kancheli,
  Yad.\ Fiz.\  {\bf 18} (1973) 595
  [Sov.\ J.\ Nucl.\ Phys.\  {\bf 18} (1974) 308].

\bibitem{Albajar:1989an}
  C.~Albajar {\it et al.}  [UA1 Collaboration],
  Nucl.\ Phys.\  B {\bf 335} (1990) 261.



\end{thebibliography}
\end{document}